\documentclass[prb,aps,superscriptaddress,showpacs,twocolumn]{revtex4-1}
\usepackage{lipsum}
\makeatletter
\newcommand*{\balancecolsandclearpage}{%
  \close@column@grid
  \clearpage
  \twocolumngrid
}
\makeatother
\usepackage[T1]{fontenc}
\usepackage{array,amsmath,amsfonts,amssymb,dsfont}
\usepackage{natbib}
\usepackage{ae}
\usepackage{color}
\usepackage{graphicx}
\usepackage{booktabs}
\usepackage{multirow}

\DeclareMathOperator{\tr}{tr}
\DeclareMathOperator{\sgn}{sgn}

\begin{document}

\title{Measurement of geometric dephasing using a superconducting qubit}

\author{S.~Berger}
\affiliation{Department of Physics, ETH Zurich, CH-8093 Zurich, Switzerland}
\author{M.~Pechal}
\affiliation{Department of Physics, ETH Zurich, CH-8093 Zurich, Switzerland}
\author{P.~Kurpiers}
\affiliation{Department of Physics, ETH Zurich, CH-8093 Zurich, Switzerland}
\author{A.A.~Abdumalikov}
\affiliation{Department of Physics, ETH Zurich, CH-8093 Zurich, Switzerland}
\author{C.~Eichler}
\altaffiliation[Current address: ]{Department of Physics, Princeton University, Princeton, New Jersey 08544, USA}
\affiliation{Department of Physics, ETH Zurich, CH-8093 Zurich, Switzerland}
\author{J.~A.~Mlynek}
\affiliation{Department of Physics, ETH Zurich, CH-8093 Zurich, Switzerland}
\author{A.~Shnirman}
\affiliation{Institut f\"ur Theorie der Kondensierten Materie,
Karlsruhe Institute of Technology, 76128 Karlsruhe, Germany}
\author{Yuval Gefen}
\affiliation{Department of Condensed Matter Physics, The Weizmann Institute of Science, Rehovot 76100, Israel}
\author{A.~Wallraff}
\affiliation{Department of Physics, ETH Zurich, CH-8093 Zurich, Switzerland}
\author{S.~Filipp}
\email[]{sfilipp@us.ibm.com}
\altaffiliation[current address: ]{IBM T.J.~Watson Research Center, Yorktown Heights, New York 10598, USA}
\affiliation{Department of Physics, ETH Zurich, CH-8093 Zurich, Switzerland}


\def\1{\mathchoice{\rm 1\mskip-4.2mu l}{\rm 1\mskip-4.2mu l}{\rm 1\mskip-4.6mu l}{\rm 1\mskip-5.2mu l}}
\newcommand{\ket}[1]{|#1\rangle}
\newcommand{\bra}[1]{\langle #1|}
\newcommand{\eval}[1]{\langle #1\rangle}
\newcommand{\braket}[2]{\langle #1|#2\rangle}
\newcommand{\ketbra}[2]{|#1\rangle\langle#2|}
\newcommand{\opelem}[3]{\langle #1|#2|#3\rangle}
\newcommand{\projection}[1]{|#1\rangle\langle#1|}
\newcommand{\scalar}[1]{\langle #1|#1\rangle}
\newcommand{\op}[1]{#1}
\newcommand{\vect}[1]{\boldsymbol{#1}}
\newcommand{\id}{\text{id}}
\newcommand{\red}[1]{\textcolor{red}{#1} }
\newcommand{\coh}[3]{\ensuremath{(#1,#2,#3)}}
\newcommand{\gam}[3]{\ensuremath{\gamma([#1,#2],#3)}}

\newcommand{\sx}{\ensuremath{\op{\sigma_x}}}
\newcommand{\sy}{\ensuremath{\op{\sigma_y}}}
\newcommand{\sz}{\ensuremath{\op{\sigma_z}}}

\newcommand{\rad}{\ensuremath{\, \mathrm{rad}}}
\newcommand{\khz}{\ensuremath{\, \mathrm{kHz}}}
\newcommand{\mhz}{\ensuremath{\, \mathrm{MHz}}}
\newcommand{\ghz}{\ensuremath{\, \mathrm{GHz}}}
\newcommand{\dbm}{\ensuremath{\, \mathrm{dBm}}}
\newcommand{\us}{\ensuremath{\, \mathrm{\mu s}}}
\newcommand{\ns}{\ensuremath{\, \mathrm{ns}}}
\newcommand{\mk}{\ensuremath{\, \mathrm{mK}}}
\newcommand{\um}{\ensuremath{\, \mathrm{\mu m}}}
\newcommand{\nm}{\ensuremath{\, \mathrm{nm}}}
\newcommand{\vpp}{\ensuremath{\, \mathrm{V_{pp}}}}

\def\be{\begin{equation}}
\def\ee{\end{equation}}
\def\bes{\begin{equation*}}
\def\ees{\end{equation*}}
\def\bea{\begin{eqnarray}}
\def\eea{\end{eqnarray}}
\def\beas{\begin{eqnarray*}}
\def\eeas{\end{eqnarray*}}

\date{\today}

\begin{abstract}
A quantum system interacting with its environment is subject to dephasing which ultimately destroys the information it holds. Using a superconducting qubit, we experimentally show that this dephasing has both dynamic and geometric origins. It is found that geometric dephasing, which is present even in the adiabatic limit and when no geometric phase is acquired, can either reduce or restore coherence depending on the orientation of the path the qubit traces out in its projective Hilbert space. It accompanies the evolution of any system in Hilbert space subjected to noise.
\end{abstract}

\maketitle

The information stored in a quantum bit is ultimately lost when the interaction with its environment causes randomization of its quantum phase in a process known as dephasing \cite{Schlosshauer2007}. Part of this dephasing is of geometric origin \cite{Whitney2005} and is related to a type of geometric phase known as Berry phase \cite{Pancharatnam1956,Berry84}, which is accumulated when a quantum system is adiabatically steered along a closed contour in the parameter space of its Hamiltonian. This phase underlines the dynamics and thermodynamics of a broad spectrum of quantum systems as measured in spins \cite{Bitter1987,Suter1987}, systems dominated by spin-orbit-coupling \cite{Meir1989,Loss1990,Nagasawa2013}, experiments exhibiting Aharonov-Bohm phases \cite{Chambers1960}, as well as atomic \cite{Atala2013,Jotzu2014} and optical setups \cite{Tomita1986,Loredo2014}.
Here, we experimentally confirm the existence of geometric dephasing using a superconducting quantum system \cite{Koch2007} exposed to artificial noise. It is found that geometric dephasing can either reduce or restore coherence depending on the orientation of the path a qubit traces out in its projective Hilbert space. This asymmetric decoherence mechanism is expected to play a role in numerous stochastic systems exhibiting geometric phases \cite{Whitney2008,Mathew2014}.

The geometric phase roots in the structure of Hilbert space and is -- unlike the dynamic phase -- not related to the duration of a quantum process. When it comes to quantum dissipative systems, ubiquitous in our physical world, the nature of the geometric phase may be put to question: noise may screen out geometric effects and the condition for adiabaticity is not self-evident. The former effect is frequently modeled by adding non-geometric dissipative rates to the equations of motion for the density matrix, see e.g. Ref.~\onlinecite{Ellinas1989}. The latter concern is
alleviated in Ref.~\onlinecite{Whitney2003}: naively one may expect that adiabaticity implies that the rate of change involved is smaller than the excitation gap. It turns out though
that a gapless spectrum, which may be the result of coupling the system to a dissipative reservoir, does not necessarily annul the Berry phase \cite{Pancharatnam1956,Berry84}.
Moreover, dephasing and level broadening due to classical low frequency noise randomizing the Berry phase are addressed in Refs.~\onlinecite{Gaitan1998,De2003}; corrections to the Berry phase, due to both slow, classical noise and high frequency quantum noise have been discussed in Ref.~\onlinecite{Whitney2005}. In that work, the concept of geometric dephasing has been introduced. Indeed, decoherence in quantum systems stems not only from the stochastic evolution of the dynamic phase of the system's wave function (dynamic dephasing), but also from geometric effects.

While the effect of classical noise \cite{Filipp2009a,Wu2013a,Berger2013} and of a quantum bath \cite{Cucchietti2010} on the Berry phase have been studied in experiments before, the present work reports on an explicit measurement of geometric dephasing. We show that the state (Bloch) vector of a qubit evolving adiabatically in the presence of Gaussian noise contains a suppression factor
\be
\label{Eq:GeneralDephasing}
\nu=\exp\left[-D(T)\left({\cal A}  + \sgn(n)\,{\cal B}\omega_B + {\cal C}\omega_B^2 +...\right)\right]\ ,
\ee
where $T \equiv 2\pi |n|/\omega_B$ is the duration of the time evolution,  $n\in\mathbb{Z}$ the oriented number of loops of the qubit in its projective Hilbert space (equal to the number of loops performed by the magnetic field for adiabatic evolution), $\omega_B>0$ the precession frequency, and $2\pi /\omega_B$ the duration of a single loop. The function $D(T)$ describes the spectral properties of the noise.
The first term in Eq.~(\ref{Eq:GeneralDephasing}) is independent of $\omega_B$ and represents dynamic dephasing. The second term, proportional to $\sgn(n)\omega_B$, represents geometric dephasing. The third term goes as $\omega_\mathrm{B}^2$ and stands for non-geometric non-adiabatic dephasing. It is especially noteworthy that the second term can either increase or decrease the total dephasing, depending on the sign of $n$. This makes the geometric nature of this term explicit. The contribution to dephasing stemming from the first term is independent of the sign of $n$ and is therefore not geometric. Likewise, the third term does not describe geometric dephasing. However, it leads to fluctuations in the Berry phase \cite{De2003} observed in Refs.~\onlinecite{Filipp2009a,Berger2013}. 

\begin{figure}
  \centering
  \includegraphics[width=80mm]{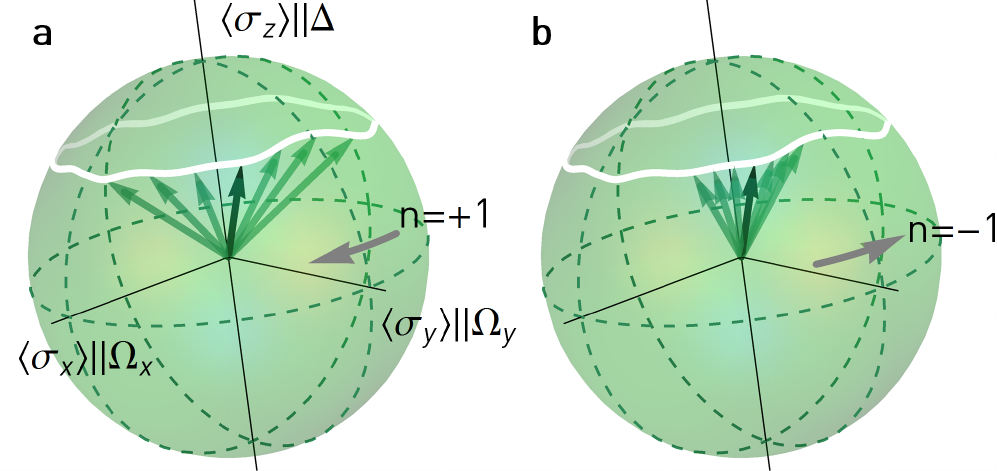}
  \caption{
\textbf{Qubit in a noisy environment.} Bloch vectors (green arrows) describing the qubit state after evolving adiabatically in an precessing noisy magnetic field. When the qubit state evolves adiabatically, its Hilbert space can be identified with the parameter space of the Hamiltonian, i.e.~the three-dimensional effective magnetic field. The magnetic field (white line) follows a closed loop $n$ times. In \textbf{a}, the number of loops is $n=+1$, in \textbf{b} it is $n=-1$. Due to geometric dephasing, the Bloch vectors are fanned out more in \textbf{a} than in \textbf{b}.
  }
  \label{fig:ill}
\end{figure}

\begin{figure}
  \centering
  \includegraphics[width=87mm]{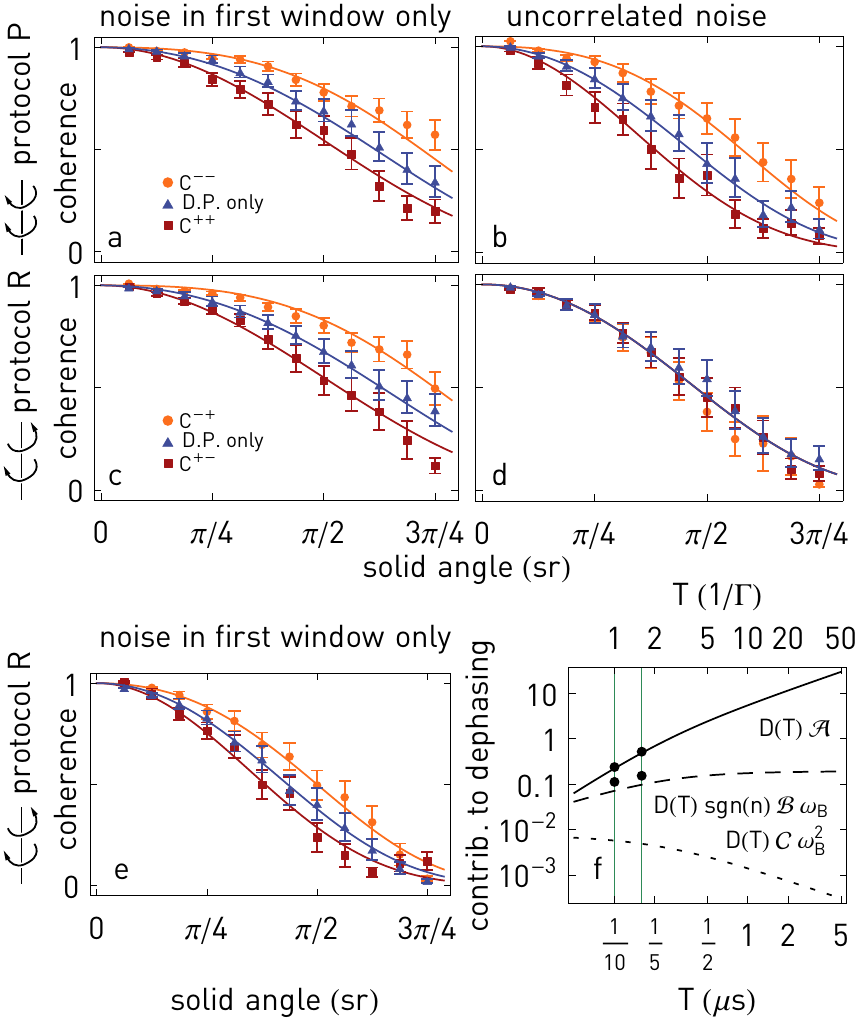}
\caption{
\textbf{Coherence of the qubit.}
Measured normalized coherences as a function of solid angle with fluctuations in the effective magnetic field. Each data point is the average of 400 realizations of noise. If not stated otherwise, the precession period of the field is $T=100$ ns. Solid lines indicate fits to the data. \textbf{a} Coherence measured in a complete spin echo experiment (`protocol P', see Fig.~\ref{fig:seq}) with noise during the first window of the spin-echo only. The direction of precession is \emph{preserved} in the second window of the echo, leading to pulse sequences $C^{++}$ and $C^{--}$. The exponents denote $\sgn(n)$, the direction of precession of the field. `D.P. only' denotes a magnetic field which does not precess ($\omega_B=0$), so that the qubit acquires only dynamic phase. \textbf{b} As in \textbf{a}, but with uncorrelated noise. \textbf{c} Coherence measured in a spin-echo in which the direction of precession of the magnetic field is \emph{reversed} in the second half  (`protocol R'), giving $C^{+-}$ and $C^{-+}$. \textbf{d} As in \textbf{c}, but with uncorrelated noise. \textbf{e} As in \textbf{c}, but with precession period $T=160$ ns. \textbf{f} Dynamic (solid line), geometric (dashed line) and non-geometric non-adiabatic (dotted line) contributions to dephasing, as they appear in the exponent of $\nu$ in Eq.~(\ref{Eq:GeneralDephasing}), as a function of precession period $T$. The curves are computed with the parameters used for recording the data shown in \textbf{c} and \textbf{e} for $A=\pi/2$. The values for dynamic and geometric dephasing extracted from fits to these measurements are indicated by black dots. The vertical lines indicate the periods $T=0.1,0.16\us$ used in our experiments.
}
  \label{fig:coh}
\end{figure}

In our experiments, a superconducting qubit (see Supplementary Information) is subject to resonant driving with slowly modulated amplitude and phase. In a frame corotating with the drive, the system can be modelled as a spin one-half particle in a slowly varying magnetic field---a paradigmatic system to observe the Berry phase \cite{Bitter1987,Leek2007,Filipp2009a}. To study geometric dephasing, we add artificial noise to the drive, mimicking noise in the magnetic field. Our analysis shows explicitly that the decay of a pure quantum state into a mixed state includes a term which is exponential in $n$, the number of oriented loops: the dephasing is asymmetric in the direction of the loops, as schematically represented in Fig.~\ref{fig:ill}. The coherence (i.e.~the length of the Bloch vector) of the qubit recorded in a Ramsey-type interferometric experiment  illustrates this phenomenon. As shown in Fig.~\ref{fig:coh}a, a cyclic change of the effective magnetic field in clockwise ($C^{++}$) or counter-clockwise ($C^{--}$) direction leads either to a decrease or an increase in coherence when compared to a static but still fluctuating magnetic field, as discussed in detail below.

The dynamics of the system in a frame rotating at the frequency $\omega_d$ of the drive are described by the Hamiltonian ($\hbar=1$) \cite{Leek2007}
\begin{equation}\label{eq:rwa}
H_1 =\frac{1}{2}\left[\Delta\,\sz + \Omega\,\cos(\varphi)\,\sx +  \Omega\,\sin(\varphi)\,\sy\right] = \frac{1}{2} \vect{B} \cdot \vect{\sigma} \ ,
\end{equation}
where $\op{\vect{\sigma}}=(\sx,\sy,\sz)$ are the Pauli matrices, $\Delta \equiv \omega_{01}-\omega_d$ the detuning in frequency between drive and qubit transition, $\Omega(t)$ the amplitude and $\varphi(t)$ the phase of the drive, and $\vect B = (\Omega\,\cos(\varphi),\Omega\,\sin(\varphi),\Delta)$ the effective magnetic field in units of angular frequency. We let the magnetic field form an angle $\theta=\arctan \Omega/\Delta$ with the $\Delta$-axis and have it precess about this axis at a rate $|\dot{\varphi}|=\omega_\mathrm{B}=2\pi|n|/T$. After a time $T$, the magnetic field has traced out $n$ loops and enclosed a solid angle $A=2n\pi(1-\cos\theta)$ as seen from $\vect{B}=0$. If $\omega_\mathrm{B}$ is small enough, that is, if the evolution is adiabatic, the qubit's state vector acquires a geometric phase $A/2$ (Ref.~\onlinecite{Berry84}). We induce fluctuations $\delta\Omega$ in the effective magnetic field in radial direction, which in our setup correspond to amplitude noise in the signal driving the qubit, so that $\Omega(t) = \Omega_0 + \delta\Omega$. Applying a transformation $H_2 = R H_1 R^{\dag} + i \dot R R^{\dag}$ with
$R = \exp\left[{i\,\varphi\,\sigma_z/2}\right]$ (see e.g.~Ref.~\onlinecite{Lax1982}) to Eq.~(\ref{eq:rwa}) results in the Hamiltonian
\be
H_2 =\frac{1}{2}\,\left[\Delta-\sgn(n)\,\omega_B \right]\,\sigma_z + \frac{1}{2}\,\left(\Omega_0+\delta\Omega\right)\,\sigma_x \ .
\label{eq:rwabis}
\ee

From Eq.~(\ref{eq:rwabis}), it follows that in a Ramsey experiment, in which the effective magnetic field performs $n$ oriented loops in the time interval $[0,T]$, the eigenstates of the qubit acquire a total relative phase $\gam{0}{T}{n}$, where
\be\label{eq:gammaTiTf}
\gam{T_i}{T_f}{n}\equiv\int_{T_i}^{T_f} dt
\sqrt{\left[\Delta - \sgn(n)\omega_B\right]^2+[\Omega_0+\delta\Omega(t)]^2}\, .
\ee
We make sure that the conditions $\delta\Omega\ll \Omega$ and $\omega_B\ll \sqrt{\Delta^2+\Omega_0^2}$ are met, so that the qubit evolves adiabatically (also see Methods section). A first-order Taylor expansion of Eq.~(\ref{eq:gammaTiTf}) in the two small parameters $\delta\Omega$ and $\omega_B$ yields
\bea\label{eq:gammaSingleRotation}
\gam{0}{T}{n} = \int_{0}^{T} dt \bigg[ \bigg.
\sqrt{\Delta^2+\Omega_0^2} -\sgn(n)\,\omega_B\cos\theta \nonumber \\
\qquad - \delta\Omega\sin\theta + \delta\Omega\frac{\sgn(n)\,\omega_B}{\sqrt{\Delta^2+\Omega_0^2}} \cos\theta \sin\theta +\ldots
\bigg.\bigg]\ .
\eea
The first two terms in Eq.~(\ref{eq:gammaSingleRotation}) give rise to $\eval{\gamma}$, the sum of dynamic and geometric phase in absence of noise. The last two terms lead to dephasing, as seen by computing the variance $\eval{(\delta\gamma)^2}$ of the phase in Eq.~(\ref{eq:gammaSingleRotation}),
\be\label{eq:variance}
\eval{(\delta\gamma)^2} =
2 D(T)\left({\cal A}  + \sgn(n)\,{\cal B}\omega_B + {\cal C}\omega_B^2 +...\right)\ .
\ee
Thus, we recover the suppression factor from Eq.~(\ref{Eq:GeneralDephasing}) 
describing the length $\nu=\sqrt{\eval{\sx}^2+\eval{\sy}^2+\eval{\sz}^2}$ of the state vector of the qubit. The function
\be
\label{eq:timecorrelator}
D(T) \equiv \frac{1}{2}\,\int_0^T d t_1\int_0^T d t_2 \eval{\delta\Omega(t_1)\delta\Omega(t_2)}
\ee
is the integrated time-correlator of the noise. The functions
\begin{eqnarray}\label{eq:ABCfactors}
{\cal A}&=&a(\sin\theta)^2\ ,\quad
{\cal B}=b\frac{2\cos\theta(\sin\theta)^2}{\sqrt{\Delta^2+\Omega_0^2}}\ ,\nonumber\\
{\cal C}&=&c\frac{(\cos\theta\sin\theta)^2}{\Delta^2+\Omega_0^2}
\end{eqnarray}
appear in the three first terms in Eq.~(\ref{eq:variance}), which represent dynamic dephasing, geometric dephasing, and non-geometric non-adiabatic corrections, respectively.

In Eq.~(\ref{eq:ABCfactors}) we have introduced dimensionless 
 decoherence factors $a$, $b$, $c$ (all equal to one in the Ramsey experiment considered) to accommodate different types of experiments, as detailed below. In particular, we use spin-echo techniques to observe geometric dephasing while eliminating the dynamic phase and enhancing the qubit coherence time.

Two protocols leading to different decoherence factors $a,b,c$ are considered. The first one is a complete echo, in which the orientation of the loop the magnetic field traverses is identical in both halves. In this protocol (named `protocol P', as in `preserved'), the phase acquired by the qubit is
\be\label{eq:phaseP}
\gamma_P=\gam{0}{T}{+n} - \gam{T}{2T}{+n}\ .
\ee

\begin{figure}
  \centering
  \includegraphics[width=87mm]{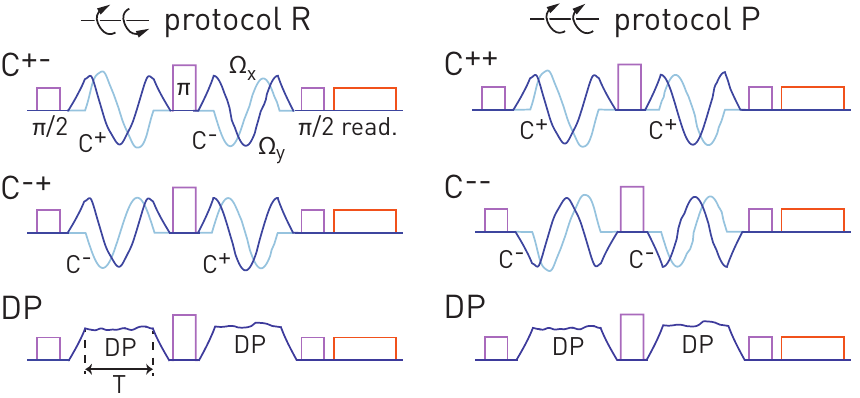}
  \caption{
  \textbf{Pulse sequences.}  Pulse sequences for protocol R (where the direction of precession of the magnetic field is reversed in the second half of the spin-echo) and protocol P (where it is preserved). The $\pi$- and $\pi/2$-pulses implementing the spin-echo are on resonance with the qubit transition frequency $\omega_{01}$ (purple). At the end of the sequence, the state of the qubit is read out by applying a tone at frequency $\omega_r$ (orange). The components $\Omega_x$ and $\Omega_y$ of the magnetic field are shown in dark and light blue. $\Delta$ is kept constant and is not shown.
}
\label{fig:seq}
\end{figure}

In the second protocol, the orientation of the loop is reversed in the second half of the echo sequence. We therefore call it `protocol R' (for `reversed'). The accumulated phase is 
\be\label{eq:phaseR}
\gamma_R=\gam{0}{T}{+n} - \gam{T}{2T}{-n}\ .
\ee
To illustrate the protocols, schematics of the pulse sequences are shown in Fig.~\ref{fig:seq}.


In our experiment, phase and coherence of the qubit have been recorded as a function of the solid angle enclosed by the effective magnetic field $\vect{B}$. 
The detuning of the off-resonant pulses is $\Delta/2\pi=-35\mhz$. The fluctuations $\delta\Omega$ applied to $\vect{B}$ conform to an Ornstein-Uhlenbeck process with correlation time $1/\Gamma=10\mhz$, intensity $\sigma^2$ and normalized noise amplitude $\sigma^2/\Omega_0^2=0.1$. Intrinsic dephasing due to a finite decoherence rate of the qubit causes the coherence to drop to about 0.7 at the end of the pulse sequence. This effect is calibrated out~\cite{Berger2013}. The correlation function of the noise process is $\eval{\delta\Omega_1(t_1)\delta\Omega_1(t_2)}=\sigma^2 e^{-\Gamma |t_1-t_2|}$, and the integrated time-correlator from Eq.~(\ref{eq:timecorrelator}) is $D(T)= \sigma^2 \left(\Gamma T-1+e^{-\Gamma T}\right)/\Gamma^2$. Since the noise $\delta\Omega(t)$ is artificial, we may control its time correlations, in particular those between the first time window, $0<t<T$, and the second time window, $T<t<2T$. For convenience, we define $\delta\Omega_1(t)\equiv \delta\Omega(t)$ and $\delta\Omega_2(t)\equiv \delta\Omega(t+T)$.

\begin{table*}
\centering
\caption{\textbf{
Decoherence factors.}
Decoherence factors for spin-echo experiments with noise on the effective magnetic field. Two different protocols are considered, with four types of noise correlations between the first and the second halves of the echo sequence (cf.~text). Theoretical decoherence factors (see Eq.~(\ref{eq:ABCfactors})) associated with dynamic dephasing ($a$), geometric dephasing ($b$), and non-geometric corrections originating from the stochasticity of the Berry phase ($c$).
}
\begin{tabular}{@{}lclllclllcllcll@{}}
\toprule
&& \multicolumn{7}{c}{theory}
&& \multicolumn{5}{c}{experiment}
\\  \cmidrule{3-9} \cmidrule{11-15}
& \phantom{a} & \multicolumn{3}{c}{prot.~R}
& \phantom{a} & \multicolumn{3}{c}{prot.~P}
& \phantom{a} & \multicolumn{2}{c}{prot.~R}
& \phantom{a} & \multicolumn{2}{c}{prot.~P}
\\  \cmidrule{3-5} \cmidrule{7-9}  \cmidrule{11-12}  \cmidrule{14-15}
correlation of noise
   && $a\phantom{a}$ & $b$\phantom{a} & $c$\phantom{a}
   && $a\phantom{a}$ & $b$\phantom{a} & $c$\phantom{a}
   && $a$ & $b$
   && $a$ & $b$
\\  \midrule
correlated, $\delta\Omega_1=\delta\Omega_2$
   &&  0  & 0  & 4
   &&  0  & 0  & 0
   &&  --- & ---
   &&  --- & --- \\
anticorrelated,  $\delta\Omega_1=-\delta\Omega_2$
   && 4  &  0  & 0
   && 4  &  4  & 4
   &&  $4.29\pm0.07$ & ---
   &&  $4.54\pm0.06$ & $5.84\pm0.20$  \\
uncorrelated, $\eval{\delta\Omega_1\delta\Omega_2}=0$
   && 2  &  0  & 2
   && 2  &  2  & 2
   &&  $2.14\pm0.04$ & ---
   &&  $2.27\pm0.03$ & $2.92\pm0.10$ \\
1st window, $\delta\Omega_2=0$
   && 1  &  1  & 1
   && 1  &  1  & 1
   &&  $1.07\pm0.02$ & $1.58\pm0.12$
   &&  $1.14\pm0.02$ & $1.46\pm0.05$ \\
\bottomrule
\end{tabular}
\label{tbl:alphabetagamma}
\end{table*}


A spin-echo sequence with noise in the first window only, i.e.~$\delta\Omega_2(t)\equiv0$, allows us to measure the dephasing the qubit experiences during a Ramsey experiment. There is geometric dephasing in both protocols ($b \neq 0$ in Fig.~\ref{fig:coh}a,c,e) which, depending on the sign of $n$, either increases or reduces the total dephasing. Interestingly, in protocol P geometric dephasing is present although the Berry phase (not shown) is eliminated along with the dynamic phase. This phenomenon can be explained by computing the variance of the phases in Eqs.~(\ref{eq:phaseP},\ref{eq:phaseR}), from which the decoherence factors $a=1$, $b=1$ follow. In our experiment, a fit to the data yields $1.14\pm0.02$ for protocol P and $a=1.07\pm 0.02$ for protocol R (Tab.~\ref{tbl:alphabetagamma}), in good agreement with computations. The observed geometric dephasing ($b=1.46\pm0.05$ for protocol P and $1.58\pm0.12$ for protocol R) is somewhat larger than predicted.
The procedure used to extract $a$ and $b$ from fits to the data is described in the Methods section.
The measured phases in protocol R agree with the prediction for a weakly anharmonic multi-level system~\cite{Berger2012}, with and without applied noise. However, when the noise causes the coherence to drop below $\approx0.2$, the phase can no longer be determined reliably and the measured phase deviates from theory.

Dynamic dephasing is more pronounced if the sequence is longer ($T=160\ns$ in panel (e), vs. $T=100\ns$ in panel (c)), but geometric dephasing is always present. The time-dependence in Eq.~(\ref{Eq:GeneralDephasing}) describes the experimental data in Fig.~\ref{fig:coh} well, with $a=1.12\pm0.03$ and $b=1.45\pm0.15$ (for $T=160\ns$) and $a=1.14\pm0.02$ and $b=1.58\pm0.12$ (for $T=100\ns$).
Examining the three contributions to dephasing appearing in the coherence suppression factor in Eq.~(\ref{Eq:GeneralDephasing})
as a function of $T$ (Fig.~\ref{fig:coh}f), we see that the dynamic contribution ($\propto{\cal A}$) grows with $T$, while the geometric contribution ($\propto{\cal B}$) saturates in the limit of short correlation times $\Gamma T\gg1$: it does \emph{not} vanish in the adiabatic limit $T\to\infty$.
The non-geometric non-adiabatic contribution ($\propto{\cal C}$) vanishes in this limit. This holds for a general noise process (see Supplementary Information).


While the experiment with noise in the first half of the spin-echo only is adequate to measure geometric dephasing, a more physically relevant scenario is to consider noise appearing in both halves of the spin-echo.

For uncorrelated noise, i.e.~when the correlation time of the noise is shorter than the timescale of the spin-echo (as e.g.~for white noise \cite{Ithier2005,Paladino2014}), we have $\langle \delta\Omega_1(t_1)\delta\Omega_2(t_2)\rangle=0$. In addition, the spectral power of the noise is kept constant, $\langle \delta\Omega_1(t_1)\delta\Omega_1(t_2)\rangle=\langle \delta\Omega_2(t_1)\delta\Omega_2(t_2)\rangle$. In this case, we measure geometric dephasing in the complete spin echo only (protocol P, Fig.~\ref{fig:coh}b and Tab.~\ref{tbl:alphabetagamma}). In protocol R, there is only dynamic dephasing (Fig.~\ref{fig:coh}d, see Methods section for details).


For perfectly correlated noise $\delta\Omega_1(t)=\delta\Omega_2(t)$, which corresponds to fluctuations slower than typical spin echo time scales as typical for $1/f$-noise, neither dynamic nor geometric dephasing is expected. Non-adiabatic contributions (quantified by $c$) only play a role in protocol R, where $c=4$ is expected and $c=2.25\pm0.25$ is found from a fit to data (not shown). In protocol P, all fluctuations cancel out ($a=b=c=0$ in theory) and a fit gives $c=0.24\pm0.54$. The effect of correlated noise has been studied in Refs.~\onlinecite{De2003,Filipp2009a,Berger2013}. 


Finally, we have also considered anticorrelated noise ($\delta\Omega_1(t)=-\delta\Omega_2(t)$) which maximizes both dynamic and geometric dephasing (Tab.~\ref{tbl:alphabetagamma}) in accordance with expectations. As with uncorrelated noise, geometric dephasing is only present in protocol P.

Although in this experiment we consider geometric dephasing for closed loops, this effect may be detected even for open trajectories in parameter space: In any interferometric setup, interference fringes originating from the non-unity overlap between the initial reference state and the final state after a non-cyclic evolution will exhibit dephasing in the form of a decaying envelope. Geometric dephasing will reveal itself as a rate of decay which depends on the direction, i.e.~the sign of $n$.

Given the broad spectrum of systems whose dynamics involves geometric phases, the presence of geometric dephasing is expected to be commonplace. While this additional contribution to decoherence is typically small, it can be of relevance for high-fidelity quantum operations and decoupling pulse techniques. For example, residual coupling to spurious modes will cause the system of interest to precess and induce dephasing with both dynamical and geometric contributions. Moreover, as shown here, geometric dephasing is present even when no geometric phase is acquired.
One resulting intriguing question concerns the effect of geometric dephasing on the braiding phase of topological quasi-particles, when the separation of the relevant particles is within reach of stochastic fluctuations of the braiding path.

\section*{Acknowledgements}
We thank Simone Gasparinetti for discussions. This work was supported by the Swiss National Science Foundation (SNF, Project 150046), the German-Israeli Foundations (GIF), the Israel Science Foundation (ISF), Minerva, and the German Science Foundation (DFG SH 81/3-1, DFG RO 2247/8-1).

\section*{Methods}
\subsection{Considerations about adiabaticity}
\label{sec:adiabaticity}

The adiabaticity parameter is defined according to Ref.~\onlinecite{Solinas2010}. Given a Hamiltonian $H(t)$, we can define an instantaneous (adiabatic) basis such that $H(t)\ket{\psi_n(t)}=E_n(t)\ket{\psi_n(t)}$. Writing D(t) for the transformation from a fixed basis (given e.g.~by H(0)) to the instantaneous basis, the Hamiltonian in the instantaneous basis is $D^{-1}(t)H(t)D(t)+\hbar w$, with $w=-iD^{-1}(t)\dot{D}(t)$. In the adiabatic case, $w$ vanishes. The adiabatic parameter is defined as $s(t)=\hbar\|w(t)\|/G$, where $\|w(t)\|=\tr\sqrt{w^\dagger(t) w(t)}$ is the trace norm of $w(t)$ and $G=\sqrt{\Omega_0^2+\Delta^2}$ is the energy gap in the spectrum of  $H(0)$. Evolution is adiabatic if $s\ll1$.

\begin{figure}
  \centering
  \includegraphics[width=87mm]{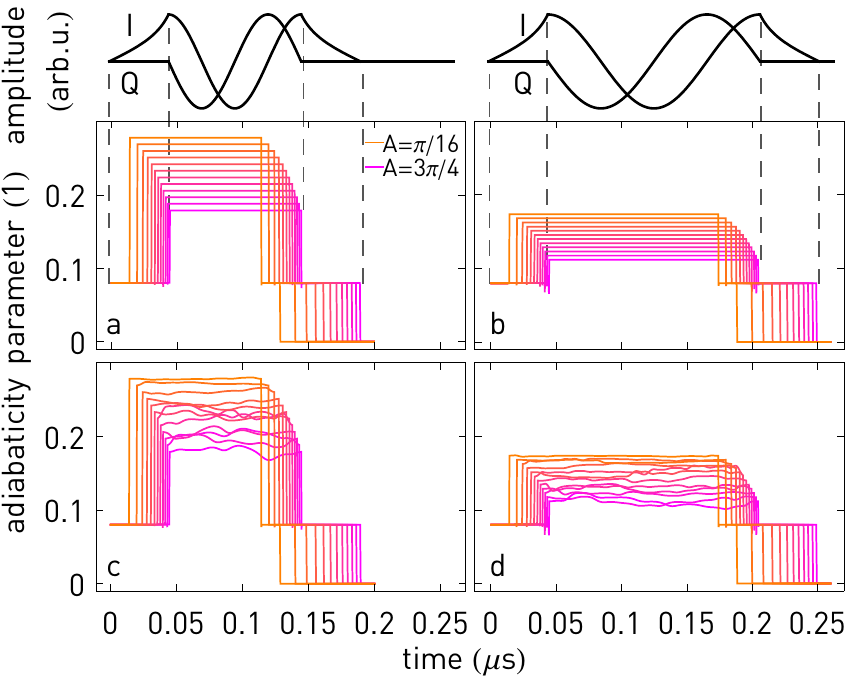}
  \caption{
    \textbf{Adiabaticity parameter.} Adiabaticity parameter during the offresonant part of the pulse sequence for $\omega_B/2\pi=T=100\ns$ (\textbf{a}) and $160\ns$ (\textbf{b}), without applied noise. Graphs for an example of a pulse sequence with applied noise are shown in panels \textbf{c} and \textbf{d}. In each panel, the adiabaticity parameter is shown for solid angles $A=\pi/16$ (orange) to $3\pi/4$ (pink) in steps of $\pi/16$. The pulse envelopes for $A=3\pi/4$ are shown on top of the plots. Note that the curves are not offset in time; the pulses have different lengths.
    }
  \label{fig:adiab}
\end{figure}

The off-resonant pulses are shaped such that $s$ is constant over time and independent of the solid angle when the drive $\Omega$ is increased or decreased (that is, at beginning and end of the pulses). For larger solid angles, the pulses need to be longer to keep $s$ constant across solid angles. When the effective magnetic field precesses (central part of the pulses), $s$ varies from solid angle to solid angle because $\omega_B$ is kept constant. The data in Fig.~\ref{fig:adiab} show that $s(t)$ is always smaller than $0.28$ and thus adiabaticity is maintained during the whole off-resonant pulse-sequence, even when noise is applied. As expected, $s$ is smaller for $\omega_B/2\pi|n|=T=160\ns$ than for $T=100\ns$.

Going to the extreme adiabatic limit $s\to0$ is not desirable. Although geometric dephasing is still present in this limit, it cannot be resolved experimentally because dynamic dephasing increases linearly with time (see Supplementary Information).

\subsection{Fitting the measured coherences}
\label{sec:fitting}

\begin{table*}[tp]
\centering
\caption{
Fitting functions used to obtain the fit estimates for $a$ and $b$ presented in Tab.~\ref{tbl:alphabetafromfits}. The column headings are abbreviations for the correlations of the noise, viz.~correlated (C), anticorrelated (A), uncorrelated (U), and noise in the first window (1). Reading example: When the fit is unconstrained, the coherences for protocol R with correlated noise and $C^{+-}$ are fitted using the function $\nu$ from Eq.~(\ref{Eq:GeneralDephasing}) with coefficients $a\rightarrow0$, $b\rightarrow-b'$ and $c\rightarrow4c$.
  }
{\footnotesize
\begin{tabular}{@{}lclllclllclllclll@{}}
\toprule
&& \multicolumn{7}{c}{unconstrained fit}
&& \multicolumn{7}{c}{constrained fit}
\\  \cmidrule{3-9} \cmidrule{11-17}
& \phantom{i} & \multicolumn{3}{c}{protocol R}
& \phantom{i} & \multicolumn{3}{c}{protocol P}
& \phantom{i} & \multicolumn{3}{c}{protocol R}
& \phantom{i} & \multicolumn{3}{c}{protocol P}
\\  \cmidrule{3-5} \cmidrule{7-9}  \cmidrule{11-13}  \cmidrule{15-17}

   && $C^{+-}$ & D.P. & $C^{-+}$
   && $C^{++}$ & D.P. & $C^{--}$
   && $C^{+-}$ & D.P. & $C^{-+}$
   && $C^{++}$ & D.P. & $C^{--}$
\\  \midrule
C
   && \coh{0}{-b'}{4c} & \coh{0}{b''}{0} & \coh{0}{b'}{4c}
   && \coh{0}{b'}{c'} & \coh{0}{b'}{c'} & \coh{0}{b'}{c'}
   && \coh{0}{0}{4c} & \coh{0}{0}{0} & \coh{0}{0}{4c}
   && \coh{0}{0}{c'} & \coh{0}{0}{c'} & \coh{0}{0}{c'}
\\
A
   && \coh{4a}{-b'}{0} & \coh{4a}{b''}{0} & \coh{4a}{b'}{0}
   && \coh{4a}{-4b}{0} & \coh{4a}{b'}{0} & \coh{4a}{4b}{0}
   && \coh{4a}{0}{0} & \coh{4a}{0}{0} & \coh{4a}{0}{0}
   && \coh{4a}{-4b}{0} & \coh{4a}{0}{0} & \coh{4a}{4b}{0}
\\
U
   && \coh{2a}{-b'}{0} & \coh{2a}{b''}{0} & \coh{2a}{b'}{0}
   && \coh{2a}{-2b}{0} & \coh{2a}{b'}{0} & \coh{2a}{2b}{0}
   && \coh{2a}{0}{0} & \coh{2a}{0}{0} & \coh{2a}{0}{0}
   && \coh{2a}{-2b}{} & \coh{2a}{0}{0} & \coh{2a}{2b}{0}
\\
1
   && \coh{a}{-b}{0} & \coh{a}{b'}{0} & \coh{a}{b}{0}
   && \coh{a}{-b}{0} & \coh{a}{b'}{0} & \coh{a}{b}{0}
   && \coh{a}{-b}{0} & \coh{a}{0}{0} & \coh{a}{b}{0}
   && \coh{a}{-b}{0} & \coh{a}{0}{0} & \coh{a}{b}{0}
\\
\bottomrule
\end{tabular}
}
\label{tab:params}
\end{table*}

\begin{table*}
\centering
\caption{Parameters estimates for $a$ and $b$ (respectively $b'$) extracted from constrained and unconstrained fits to data in Fig.~\ref{fig:coh} and additional data (not shown). For the unconstrained fit to the data of protocol R with D.P., $b''=-0.08\pm0.08$ is found.
}
\begin{tabular}{@{}lcllcllcllcll@{}}
\toprule
&& \multicolumn{5}{c}{unconstrained fit}
&& \multicolumn{5}{c}{constrained fit}
\\  \cmidrule{3-7} \cmidrule{9-13}
& \phantom{a} & \multicolumn{2}{c}{protocol R}
& \phantom{a} & \multicolumn{2}{c}{protocol P}
& \phantom{a} & \multicolumn{2}{c}{protocol R}
& \phantom{a} & \multicolumn{2}{c}{protocol P}
\\  \cmidrule{3-4} \cmidrule{6-7}  \cmidrule{9-10}  \cmidrule{12-13}
correlation of noise
   && $a\phantom{a}$ & $b, b'$\phantom{a}
   && $a\phantom{a}$ & $b, b'$\phantom{a}
   && $a$ & $b$
   && $a$ & $b$
\\  \midrule
correlated
   &&  ---  & $0.06\pm0.07$
   &&  ---  & $-0.30\pm0.12$
   &&  --- & ---
   &&  --- & --- \\
anticorrelated
   && $4.31\pm0.08$  & $0.06\pm0.07$
   && $4.63\pm0.07$  &  $\phantom{-}6.00\pm0.21$
   &&  $4.29\pm0.07$ & ---
   &&  $4.54\pm0.06$ & $5.84\pm0.20$  \\
uncorrelated
   && $2.15\pm0.04$  &  $0.06\pm0.07$
   && $2.32\pm0.04$  &  $\phantom{-}3.00\pm0.11$
   &&  $2.14\pm0.04$ & ---
   &&  $2.27\pm0.03$ & $2.92\pm0.10$ \\
1st window
   && $1.08\pm0.02$  &  $1.59\pm0.12$
   && $1.16\pm0.02$  &  $\phantom{-}1.50\pm0.05$
   &&  $1.07\pm0.02$ & $1.58\pm0.12$
   &&  $1.14\pm0.02$ & $1.46\pm0.05$ \\
\bottomrule
\end{tabular}
\label{tbl:alphabetafromfits}
\end{table*}

This section describes the fit models used to extract the parameters $a$, $b$ and $c$ quantifying dynamic and geometric dephasing.

In the fitting procedure, in a first step the effective normalized noise amplitude is found by fitting the function Eq.~(\ref{Eq:GeneralDephasing}) describing the coherence $\nu$ to the data from protocol R with noise in the first window and dynamic phase (D.P.) only, assuming $a=1, b=0,c=0$ and with the normalized noise amplitude as the only fit parameter. In this way, a normalized noise amplitude of $\sigma/\Omega_0^2=0.085$ 
was determined, which is slightly smaller than the set value 0.1.

In a second step, all data from protocol R are fitted simultaneously for coefficients $a$ and $b$ of the functions ${\cal A}$ and ${\cal B}$. Where the theory predicts $b=0$, we do not wish to constrain the fitting function (and, by extension, limit the model) by setting $b=0$. Rather, we use a fitting function with a separate, primed variable, which ideally the fit then shows to be zero. We note that using the same variable $b$ as in the other fits, where $b\neq0$ is expected, is not possible, since all data is fitted simultaneously. As an example, consider protocol R with anticorrelated noise and $C^{-+}$, where according to Tab.~\ref{tab:params} no geometric dephasing ($b=0$) is expected: A fitting function with $b\rightarrow b'$ is used and the fit ideally produces $b'=0$. Similarly, for protocol R, D.P., a variable $b''$ is introduced to avoid unwanted interference with $b'$, as using a single parameter would prevent $b'$ from taking values other than zero in the fits to protocol R, $C^{+-}$ and $C^{-+}$, where $b'=0$ is expected. Finally, the proportionality factors of $a$ and $b$ in the fitting functions are fixed across the measurements. For instance, protocol R with anticorrelated noise uses $4a$ and protocol R with uncorrelated noise uses $2a$. The data from protocol P is fitted similarly.

In a third step, the parameter $c$  is only fitted for protocol R, correlated noise, $C^{\pm}$, where $a$ and $b$ vanish. This is the only measurement where it is relevant, 

Fourth, the fitting parameters $b'$ and $b''$ are set to zero in the fit we call `constrained'. When dropping this constraint, the fitted values for $a$ and $b$ increase slightly (by maximally $6\%$, see Tab.~\ref{tbl:alphabetafromfits}); in turn, $b'$ and $b''$ become negative.
Simply put, the `unconstrained' model (where $b'$ and $b''$ are free parameters) trades off some fitting parameters against each other in order to obtain the best fit. If theory was in perfect agreement with experimental data (no noise, no systematic errors), this trade-off would not be possible. However, due experimental imperfections the fit produces an unphysical result, namely `negative' dynamic dephasing ($b',b''<0$). To avoid this effect, we have opted for presenting the constrained model in the main text. Comparing the parameter estimates in Tab.~\ref{tbl:alphabetafromfits}, it becomes apparent that both models yield similar parameter estimates. In addition, as discussed in the Supplementary Information, both models (constrained and unconstrained) have empirical support.





\balancecolsandclearpage
\begin{center}
\textbf{\large Supplementary Information: Measurement of geometric dephasing using a superconducting qubit}
\end{center}
\setcounter{equation}{0}
\setcounter{page}{1}
\newcommand{\beginsupplement}{%
        \setcounter{table}{0}
        \renewcommand{\thetable}{S\arabic{table}}%
        \setcounter{figure}{0}
        \renewcommand{\thefigure}{S\arabic{figure}}%
        \renewcommand{\theequation}{S\arabic{equation}}
     }

\beginsupplement

\section{Dependence of decoherence on the temporal correlations of the
noise}
\label{sec:shortlonglimits}

In this appendix, an arbitrary noise process is considered, characterized by its
intensity $\sigma^2$ (the integral of its power spectral density) and its
correlation time $1/\Gamma$. We show that only the second term $\propto
(\omega_B)^1$ represents geometric dephasing by estimating the coherence
suppression factor $\nu$ in two limiting cases:
noise with short correlation time ($\Gamma T\gg1$) and noise with long
correlation time ($\Gamma T\ll1$).

{\it Short correlation time.} In this case, $\Gamma T\gg1$ and therefore $D(T)
\propto \sigma^2 T/\Gamma$. Eq.~(1) in the main manuscript thus becomes
\be
\nu=\exp\left[-O(1)\frac{2\pi\sigma^2}{\Gamma}
\left(\frac{{\cal A}}{\omega_B}\,|n|
+ {\cal B}\, n  + {\cal C}\omega_B\,|n| +...\right)\right]
\ee
and we see that only the second term depends on the sign of $n$ and
represents geometric dephasing.

{\it Long correlation time.} Here $\Gamma T\ll1$, and we obtain $D(T) \propto
\sigma^2 T^2$, leading to
\begin{multline}
\nu=\exp\bigg[-O(1)(2\pi\sigma)^2\,\\
\times\left(\frac{{\cal A}}{\omega_B^2}\,|n|^2  +\frac{{\cal
B}}{\omega_B}\, |n| n  + {\cal C}\,|n|^2 +...\right)\bigg]\ .
\end{multline}
Again, only the second term contributes to geometric dephasing. To sum up,
there is a geometric contribution to dephasing regardless of the correlation
time of the noise (or, put differently, regardless of the duration of the
evolution).

\section{Description of the sample and the setup}
\label{sec:systemparams}

\begin{figure}
  \centering
  \includegraphics[width=87mm]{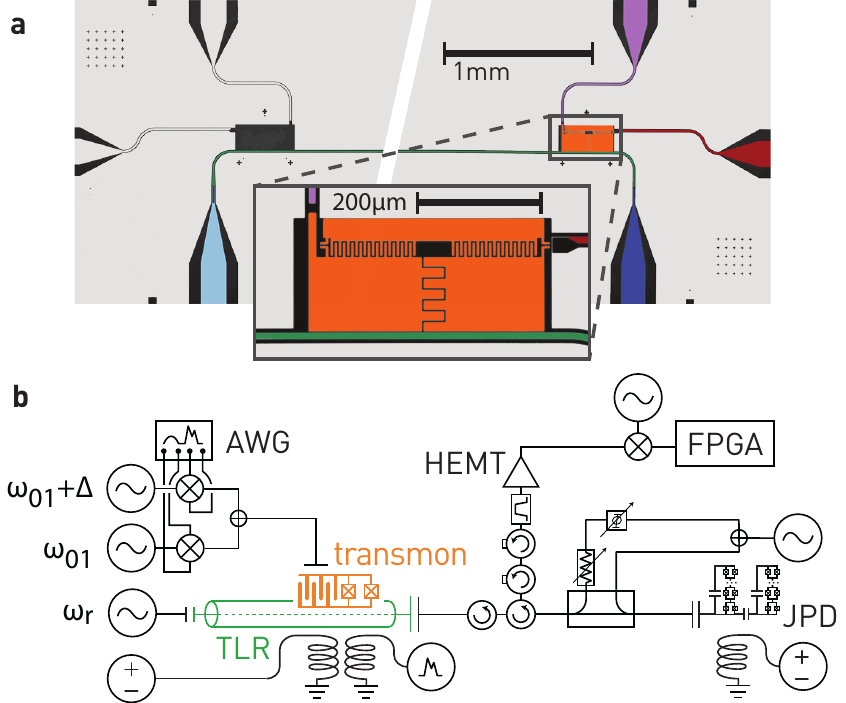}
  \caption{
(a) Micrograph of the sample. Resonator input (dark blue), resonator output
(light blue), resonator (green), charge line (red), flux line (violet) and transmon
qubit (orange), and zoom on the qubit (inset).
(b) Schematic of the measurement setup. See text for details.
  }
  \label{fig:sample}
\end{figure}

The sample (Fig.~\ref{fig:sample}) is a superconducting circuit coupled to a transmission line resonator (TLR) \cite{koch2007a}. The qubit consists of three superconducting islands (shown in orange in Fig.~\ref{fig:sample}) and two squid loops.
This device\cite{Srinivasan2011} can be seen as two coupled transmons forming a superconducting qubit with tunable frequency and tunable coupling to the resonator.
Both parameters are tuned using a static bias current applied to a flux line and a miniature coil mounted below the sample. In the experiment presented here, the bright state\cite{Srinivasan2011} serves as a qubit, whose coupling is tuned to $g/2\pi=38 \mhz$ and kept constant. Its lowest transition is set to a frequency $\omega_{01}=7.0335\ghz$. The dark state\cite{Srinivasan2011} is tuned so that its transition frequency lies above $\omega_{01}$ and thus can safely be ignored.

The qubit has an anharmonicity of $\alpha/2\pi=90\mhz$ as determined by spectroscopical measurements. It is inherently low due to the tunable-coupling design. When applying resonant pulses on the transition $\ket{0}\leftrightarrow\ket{1}$, we use pulses with truncated-gaussian envelopes of 20 ns duration. To avoid populating the higher excited states of the qubit,  a pulse-shaping technique known as derivative removal by adiabatic gate is employed \cite{Motzoi2009}.

The fundamental mode of the resonator is at a frequency
$\omega_r/2\pi=7.347\ghz$ and has a loaded quality factor of $Q=3600$. We
dispersively read out \cite{Bianchetti2009} the quantum state of the qubit by
monitoring the transmission of a rf-signal at frequency $\omega_r$ trough the
resonator. The input port of the resonator is coupled less strongly to the
transmission line than the output port to increase the SNR of the readout.

Via two circulators and a directional coupler, the output signal of the resonator
goes to an amplifier based on a Josephson parametric dimer (JPD,
Ref.~\onlinecite{Eichler2014a}), with a gain of 18.4 dB at a bandwidth of 28 MHz
centered around $7.348\ghz$. The first circulator prevents reflected signals in
the output line from leaking into the cavity; the second circulator separates
the output of the JPD amplifier from the input. The directional coupler is used
to operate the JPD amplifier: a tone is split, the first half pumps the JPD and
the second is phase-shifted and attenuated so that it cancels the pump tone
in the amplified signal. The pump tone is applied at a frequency $\omega_p=7.564\ghz$,
detuned $217\mhz$ from the resonator.

The signal is then bandpass filtered (4 to 8 GHz) and amplified with a
high-electron-mobility transistor (HEMT) providing 35 dB of gain. At room
temperature, the signal is amplified further, filtered and downconverted to 25
MHz before it is digitized at a rate of 100 MS/s with an analogue-to-digital
converter and then processed with a field-programmable gate array (FPGA).

We use direct modulation of an in-phase/quadrature mixer to generate the
microwave pulses for qubit state manipulation. The pulses are applied through
a capacitively coupled charge bias line. The I and Q quadratures are
synthesized with an arbitrary waveform generator (AWG). The pulses resonant
with the qubit transition have a gaussian envelope with a standard deviation of
$\sigma=10\ns$. They are symmetrically truncated to a length of $4\sigma$.
Measuring Rabi oscillations allows us to extract the amplitudes of $\pi$- and
$\pi/2$-pulses. We perform Ramsey interferometry experiments to calibrate the
frequency of these pulses, which also serve to extract the dephasing time
$T_2^\star=770\ns$ and echo-decay-time $T_{2}^\mathrm{echo}=1520\ns$
of the $\omega_{01}$-transition. The lifetime of the first excited state is
found to be $T_1=1330\ns$.

\section{Comparison of fitting models}

In this section, the constrained and the unconstrained fitting models are compared.

\begin{figure}
  \centering
  \includegraphics[width=86mm]{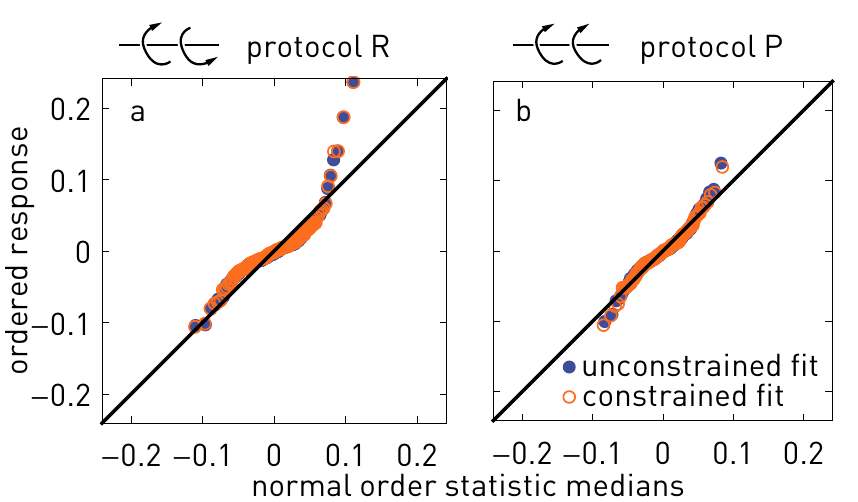}
  \caption{
Normal probability plots of the residues of the fits from protocol R and P. The
line indicates what is expected for a perfect normal distribution.
  }
  \label{fig:normalprobplot}
\end{figure}

The normal probability plots \cite{Chambers1983} of the residues of the
models, Fig.~\ref{fig:normalprobplot}, show that the residues fall near the line
describing the identity function, with the exception of a few residues which are
larger than expected. Therefore, they are normally distributed. This holds for
protocols R and P in both models (constrained and unconstrained). The
underlying normal probability distributions of the residuals assume a mean 0 and
use the standard deviation computed from the residuals:
\bea
\label{eq:stddev}
\sigma_{\mathrm{unc.}}^{(R)}=0.0425, \quad
\sigma_{\mathrm{con.}}^{(R)}=0.0427, \nonumber \\
\sigma_{\mathrm{unc.}}^{(P)}=0.0319, \quad
\sigma_{\mathrm{con.}}^{(P)}=0.0326.
\eea
%
estimation is permissible.
To estimate if constrained and unconstrained models fit similarly well, we
consider the Akaike information coefficient (AIC) \cite{Burnham2002}. The
number of fitting parameters is
\be
k_{\mathrm{unc.}}^{(R)}=6, \quad
k_{\mathrm{con.}}^{(R)}=4, \quad
k_{\mathrm{unc.}}^{(P)}=5, \quad
k_{\mathrm{con.}}^{(P)}=4,
\ee
where we take into account the fact that the variance of the residuals
Eq.~(\ref{eq:stddev}) is estimated from the model. Since the number of fitting
parameters $k$ is not very small compared to the sample size
$N=3\cdot4\cdot12=144$ (the number of measured coherences), we use
$\mathrm{AIC}_c$, the finite-sample-corrected AIC. Because the residuals are
normally distributed, and assuming the variance of the residuals is constant,
$\mathrm{AIC}_c$ takes on the simple form
\be
\mathrm{AIC}_c = N \ln \left( \mathrm{RSS} /N \right) + 2k +
\frac{2k(k+1)}{N-k-1},
\ee
with the residual sum of squares $\mathrm{RSS}$.
When comparing two models, the better model has the smaller
$\mathrm{AIC}_c$ value. As a rough rule of thumb \cite{Burnham2002}, if the
difference in $\mathrm{AIC}_c$ with the second model lies between 0 and 2,
that model has substantial empirical support, if between 4 and 7, it has less
support, and if larger than 10 it should not be considered. Here, for protocol R
the unconstrained model is the better model and the difference in
$\mathrm{AIC}_c$ with the constrained model is 2.41. For protocol P, the
converse is true: The constrained model is a better model and the difference in
$\mathrm{AIC}_c$ with the unconstrained model is 4.31. To sum up, for both
protocols the constrained model and the unconstrained one have some
empirical support.

In addition, a $t$-test has been performed to assess the significance of the
parameters $a, b, b', c, c'$ in the unconstrained model.
With a sample size of $N=144$ and $k_{\mathrm{unc.}}^{(R)}=5$,
respectively $k_{\mathrm{unc.}}^{(P)}=4$ degrees of freedom, we find (for
both) a threshold $t=1.97$ at which a parameter value has a non-zero value
with $95\%$ of significance in a two-sided $t$-test. Given that
\bea
t_{a}^{(R)}=55.53,\quad
t_{b}^{(R)}=12.79,\nonumber\\
t_{b'}^{(R)}=0.88,\quad
t_{c}^{(R)}=9.18,
\eea
and
\bea
t_{a}^{(P)}=65.22,\quad
t_{b}^{(P)}=28.14,\nonumber\\
t_{b'}^{(P)}=-2.56,\quad
t_{c'}^{(P)}=2.56,
\eea
it can be asserted that in both protocols the parameters $a$ quantifying the
dynamic dephasing and $b$ quantifying the geometric dephasing are
significant. Furthermore, the $t$-values for the parameter $b'$ are close to
zero (as it should be, since we expect $b'=0$), indicating that in this
measurement there is no geometric dephasing. Finally, the value
$t_{c}^{(R)}$ indicates significance of the parameter $c$ in protocol R,
where non-geometric non-adiabatic dephasing is present. In protocol P, where
$c'=0$ is expected, it is only weakly significant.

We note that fitting for the parameters $a$, $b$ and $c$ for individual data
sets with $n=12$ (such as protocol R with correlated noise and $C^{-+}$) or
groups of data sets with $n=36$ (such as protocol R with correlated noise and
either $C^{-+}$, $C^{+-}$ or D.P.) does not produce useful parameter
estimates. The same phenomenon as described above, the trading off of some
geometric dephasing against dynamic dephasing, is exacerbated and the fit
values are not significant.



\end{document}